\documentclass[12pt,preprint]{aastex}

\shortauthors{Gizis, Shipman and Harvin}
\shorttitle{Brown Dwarf}

\begin{document}

\title{First Ultraviolet Spectrum of a Brown Dwarf: Evidence for H$_2$ Fluorescence and Accretion\footnote{Based on observations made with the NASA/ESA Hubble Space Telescope, obtained at the Space Telescope Science Institute, which is operated by the Association of Universities for Research in Astronomy, Inc., under NASA contract NAS 5-26555. These observations are associated with program \# 9841.}
}

\author{John E. Gizis, Harry L. Shipman, James A. Harvin}
\affil{Department of Physics and Astronomy, University of Delaware, 
Newark, DE 19716}

\begin{abstract}
We analyze an HST/STIS ultraviolet spectrum of the young brown dwarf
2MASSW J1207334-393254, a member of the ten million-year old
TW Hya Association that has a planetary-mass companion.  
We detect and identify numerous emission lines.  CIV and other
ions are seen that arise in hot gas.  
We identify a series of lines with Lyman-pumped H$_2$ molecular lines,
indicating that cool gas is also present. 
Overall, this substellar object shows many of the same characteristics as 
classical T Tauri stars.  We interpret our results as direct evidence of 
accretion from a circumstellar gas disk, consistent with previous claims.     
The lack of SiIV emission from the accreting gas indicates that silicon has 
been depleted into grains.  
\end{abstract}


\section{Introduction}

It has long been speculated that the star formation process results in both stars
above and brown dwarfs below the hydrogen-burning limit.  With the discovery of 
many brown dwarfs in star-forming regions, their formation and early evolution
can now be observed.  Infrared excesses indicate that many brown dwarfs
are surrounded by dusty disks \citep{jaya,liu,mohanty_ir}.  In at least one case, a brown dwarf
of $0.015 M_\odot$, just above the deuterium-burning limit, has a disk \citep{luhman}. 
Twenty-seven out of seventy-four young ($<3$ Myr) very-low-mass stars and 
brown dwarfs also have optical emission lines including broad H$\alpha$,  which are
interpreted as due to accretion \citep{mohanty_bd}.  The accreting brown dwarfs are thus considered analogs of substellar classical T Tauri stars.  The star forming region brown dwarfs are too
distant to resolve close binaries, but Hubble Space Telescope studies of nearby older field 
brown dwarfs indicate that 20\% of them are binaries with separations
$\lesssim 10$ A.U.  \citep{hstl2}.  In summary, there is ample evidence that 
in many respects the formation of brown dwarfs is similar to that of stars.  

The closest known brown dwarf with a disk is 2MASSW J1207334-393254\footnote{This source appears as 2MASS J12073346-3932539 in the final 2MASS release.} .  Gizis (2002) discovered
this $\sim 0.03 M_\odot$  brown dwarf in a search for substellar members of the
$\sim 10$ Myr old TW Hya Association (TWA) and noted its strong H$\alpha$ emssion.  
Mohanty et al. (2003) found that the emission is variable and very broad, and argued that it
is due to accretion.    \citet{twhyax} set an upper limit to the presence of X-rays, and
argued that this showed that no more that 10\% of the H$\alpha$ emission
is chromospheric, thus supporting the accretion origin of the H$\alpha$ emission.
\citet{gemini} detected an infrared excess at  8.7 and 10.4 microns
and argued for the presence of a disk.   \citet{jaya} earlier showed there is no L' (3.8 micron) excess,
indicating the disk may have a growing inner hole as in the TWA members TW Hya and
Hen 3-600A \citep{jaya99a,jaya99b}.
It is worth noting that the system's youth is confirmed by its low
surface gravity (Gizis 2002) and lithium (Mohanty et al. 2003), and 
TWA Membership is confirmed by both radial velocity (Mohanty et al. 2003) and proper motion
(Gizis 2002, Scholz et al. 2005).  TW Hya itself is at a distance of 55 parsecs \citep{hipparcos}, but it has
been argued that 2M1207 is further away, perhaps at 70 parsecs (Sterzik et al. 2004).  
As if this system were not interesting enough, \citet{planet1} detected a candidate
giant planet companion at a separation of  0.78 arcseconds or 55 A.U..  
The L dwarf nature of the companion has been confirmed by \citep{planet2}.
Proper motion confirmation is underway.    

In this paper, we present an ultraviolet spectrum of 2M1207.  Unlike objects in 
dusty star-forming regions, the TWA is relatively free of extinction, making 2M1207 
an especially favorable target for ultraviolet spectroscopy.  

\section{Observations and Data Reduction}

We observed 2M1207 on  24-25  July 2004 using Hubble Space Telescope and the
Space Telescope Imaging Spectrograph (STIS) in spectroscopic mode.
The grating was G140L with the FUV MAMA detectors using the 0.2 arcsecond slit.  
This setup gives coverage from 1100 \AA~ to 1700 \AA.
Three exposures of 2150 seconds (starting at UT 23:24:43), 2900 seconds (UT 00:46:34) and 2900 seconds (UT 02:22:31) were obtained.  We used
the flux-calibrated two-dimensional images as created by the standard pipeline.
No continuum is detected in the spectra, but numerous spectral lines are
strongly detected.  We traced these using standard IRAF tasks and extracted the
spectra.  No significant differences were seen in the spectra extracted from
the three individual images (specifically, the CIV fluxes agree to within 8\% in the three exposures); for the remainder of this paper, we use the spectrum
extracted from an image created by averaging the three individual exposures.
This spectrum is shown in Figure 1 and the measured line fluxes are listed in Table 1.  
To check on our procedures, we used the same procedures on the three field
late-M dwarfs (VB8, VB10 and LHS 2065) observed with the same setup by \citet{hawleyuv}, and found the resulting spectra were consistent with the published ones.

\section{Analysis}
Superficially, the ultraviolet spectrum strongly resembles IUE spectra of classical T Tauri stars
\citep{valenti} while differing in many respects from the field late-M dwarfs.  Most 
strikingly, numerous H$_2$ fluorescence lines are present.  
 \citet{twhyauv1} present line identifications for the H$_2$ lines observed
in high-resolution STIS observations of TW Hya.  All of our detected 
features match lines pumped by Ly $\alpha$ 1215\AA and 1216\AA photons.
We do indeed detect strong Ly$\alpha$ directly in our spectrum, but its strength must be affected by interstellar absorption and geocoronal contamination.  the lack of spatial
extent of the emission lines indicates that the emission is from the inner
$\sim 0.2$ arcseconds (14 A.U.).    

In addition to the molecular H$_2$ emission, we detect ions that must arise from
very hot gas: CIV, CII, CIII, HeII, NV, and OI.   
All are  present in both classical T Tauri stars and field M dwarfs.  However, the
emission lines differ significantly from the field dwarfs.   
In 2M1207, the two components of the CIV double are nearly
equal strength (note that the apparently stronger blue component is blended with 
an H$_2$ line), whereas in the field dwarfs the blue component is twice as strong.  
This indicates 2M1207's CIV emission is optically thick, while the field dwarf's
transition region emission is optically thin. 
We fail to detect SiIV, which would be easily detectable if in the same ratio
as the field dwarfs, and which should arise in similar temperatures as the CIV.  
In classical T Tauri stars, this Si feature is also sometimes much weaker than expected  \citep{valenti}.  

We interpret these observations as showing that 2M1207 is actively accreting from
a circumstellar gas disk, following the scenarios already worked out for
classical T Tauri stars \citep{twhyauv2}[and references therein] and as
advocated by \citet{mohanty} for this source.  
This circumstellar gas is, for the first time in a brown dwarf, detected
directly in the form of the H$_2$ emission lines.  The accretion columns have
already been described by the observations of broad H$\alpha$ emission \citet{mohanty}.
Near the surface, this gas is (shock) heated to high temperatures ($T \approx 10^5$K).
These hotspots are responsible for the CIV emission.  The lack of Si emission in
the hotspots is due to its depletion into dust grains in the circumstellar disk.
Indeed, the dusty disk with silicate emission has been detected in the mid-infrared \citep{gemini}.  
The lack of detectable X-ray coronal emission confirms that 
chromospheric and transition region emission is relatively weak 
in 2M1207 compared to the emission from mechanisms described above
\citep{twhyax}.   Accretion is evidently variable within a night  as well as on long timescales given
the observed variability of H$\alpha$ \citep{mohanty,twhyax},  complicating any 
effort to compare non-simultaneous datasets.  Overall, 2M1207 seems to be
very similar to TW Hya itself.     Three other TWA stars, Hen 3-600A, TWA 14  and TWA 5A \citep{muze01,mohanty},  also appeared to accreting.  

These observations are particularly interesting given that 2M1207 has a giant planetary mass
companion candidate \citep{planet1,planet2}.   \citet{twhyair} argue that TW Hya originally had an unusually large circumstellar disk of $\sim 25$\% of the primary mass.  The
long accretion lifetime and observed similarity of 2M1207 raises the possibility that
perhaps 2M1207 also had an unusually large disk.  The candidate companion
would represent 10-20\% of the primary's mass; it is an suggestive coincidence that 
2M1207 might have an unusually massive disk and has an unusual planetary-mass
companion.  However, since the estimated mass of the planet would then still be 
comparable to the disk it is difficult to imagine that the companion formed like
the planets in our own Solar System.    The most likely explanation is that the system formed as
double brown dwarfs.  
The system is more widely separately than any field double brown dwarfs, although existing
searches were not sensitive to such extreme mass ratios \citep{hstl2}.     Planet formation
may also be occurring in the inner disk where we observe depleted gas.  

\section{Summary}

We have obtained the first ultraviolet spectrum of a {\it{bona fide}} brown dwarf.  We observe
CIV and other emission lines which must arise in hot gas, which we identify as
due to accretion, and H$_2$ lines which must arise in much colder gas, which
we identify as circumstellar gas.  This brown dwarf is very similar to a classical
T Tauri star.  The emission lines differ significantly from those of late-M field dwarfs,
and we therefore rule out origin in a transition region between the chromosphere and corona.

Further studies of this fascinating system are in order.  We have presented our observations in
this paper, but further modeling is required in order to estimate the accretion mass rate and other
properties of the system. We are obtaining Spitzer mid-infrared photometry of 2M1207 in order to characterize the disk.  Sub-mm observations are needed to estimate the disk mass.

\acknowledgments

We thank Jeff Valenti, Dermott Mullan, and Stan Owocki for useful conversations.  
Support for program \#9841 was provided by NASA through a grant from the Space Telescope Science Institute, which is operated by the Association of Universities for Research in Astronomy, Inc., under NASA contract NAS 5-26555.

\begin{deluxetable}{llr}
\tablewidth{0pc}
\tablenum{1}
\tablecaption{Emission Line Strengths}
\tablehead{
\colhead{Ion} &
\colhead{$\lambda$ (\AA)} &
\colhead{Flux ($10^{-16}$~erg/sec/cm$^2$)} 
}
\startdata
CIII & 1176 & 1\\
NV & 1238 & 5\\
SI & 1296 & $<1$\\
OI & 1304 & 2\\
CII & 1335 & 5\\
OI & 1356  & 1 \\
SiIV & 1394 & $<1.5$ \\
SiIV & 1403& $<1.5$\\
CIV & 1549 & 29\\
HeII & 1640 & 6\\
\enddata
\end{deluxetable}

\begin{figure}
\epsscale{0.7}
\plotone{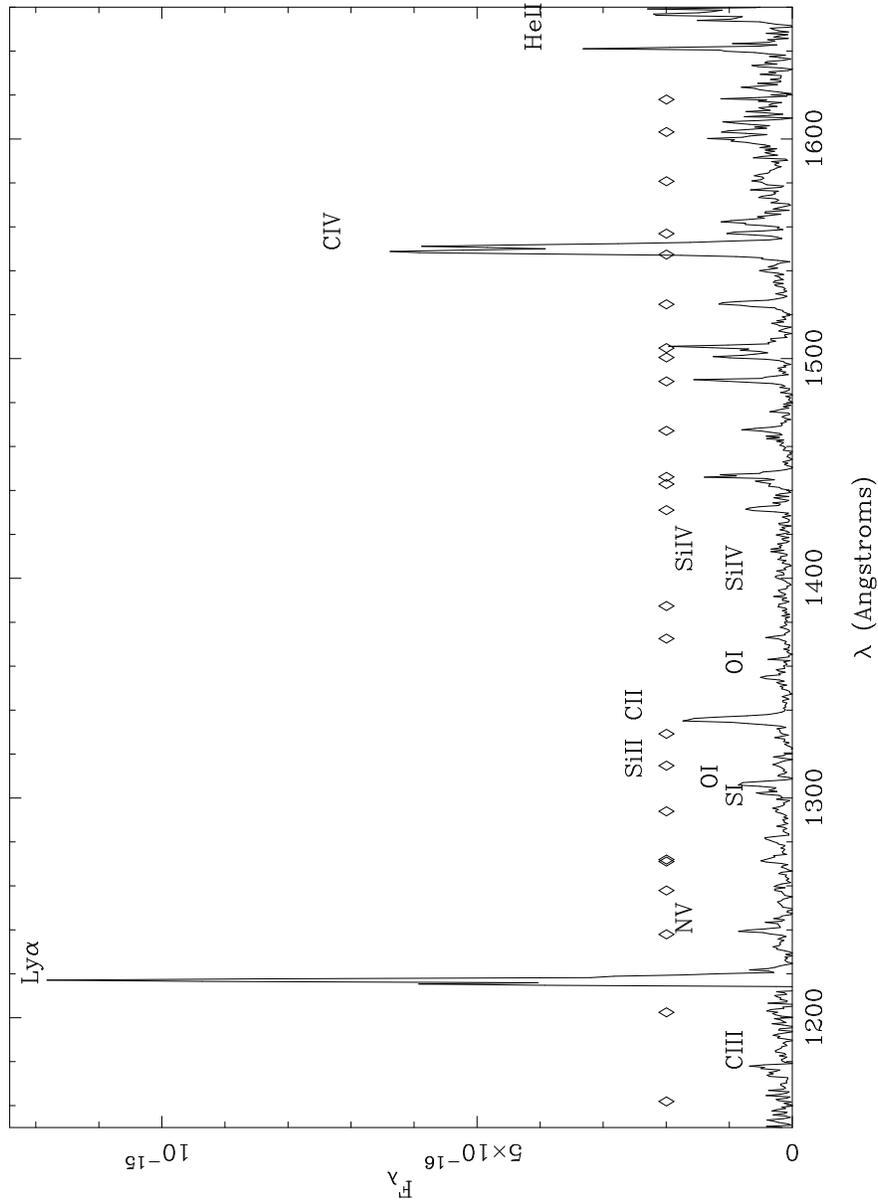}
\caption{HST/STIS spectrum of 2M1207.  We detect numerous ions.  The positions of
SiIV and SiII are marked, but they are not detected.  The diamonds mark the positions of
the H$_2$ lines pumped by  1215\AA~ and 1216\AA~ identified in TW Hya by \citet{twhyauv1}.
Most are clearly detected, and those absent are significantly weaker in the TW Hya spectrum
and thus are expected to be too faint to detect.  
\label{fig1}}
\end{figure}

\end{document}